\documentclass[a4paper,11pt]{article}
\pdfoutput=1

\usepackage{jcappub}

\usepackage{multirow}
\usepackage{subcaption}
\usepackage{enumerate}
\newcommand{\be}{\begin{equation}}
\newcommand{\ee}{\end{equation}}
\newcommand{\bea}{\begin{eqnarray}}
\newcommand{\eea}{\end{eqnarray}}

\renewcommand{\d}{\mathrm{d}}

\title{A cosmographic analysis of the transition to acceleration using SN-Ia and BAO}
\author[a]{Daniel Muthukrishna}
\author[a]{David Parkinson}

\affiliation[a]{School of Mathematics and Physics, University of Queensland, \\Brisbane, QLD 4072, Australia}
\emailAdd{d.muthukrishna@uq.edu.au, d.parkinson@uq.edu.au}

\abstract{
We explore the distance-redshift relation using a cosmographic methodology, and show how the cosmographic parameters can be used to determine the redshift of transition from deceleration to acceleration. Such a transition at a low redshift occupies only a small region of the available parameter space, and the prior assumption of an early period of deceleration can significantly change the posterior constraints. We use available type Ia Supernovae (SN-Ia) and Baryon Acoustic Oscillation (BAO) data sets to determine the cosmographic deceleration $q_0$, jerk $j_0$, snap $s_0$ and lerk $l_0$ parameters. The parameters are consistent with the $\Lambda$CDM model for a flat universe within 2-sigma. We derive constraints on the redshift of transition from deceleration to acceleration for the different expansions, and find $z_{\rm acc} > 0.14$ at 95\% confidence in the most conservative case.}

\keywords{dark energy experiments, supernova type Ia - standard candles, galaxy clustering}

\begin{document}
\maketitle
\flushbottom


\section{Introduction}

While the notion of an expanding universe has been well known for over eighty years, the 1998 discovery \cite{1998ApJ...507...46S,1998AJ....116.1009R,1999ApJ...517..565P} that this expansion rate was accelerating was a major challenge to our understanding of the composition of the universe. For this reason, it is imperative that this discovery is confirmed directly with a range of models and datasets.

Confirmation of the acceleration through a model-independent analysis of the data was suggested early on, through the statefinder approach \cite{StatefinderA,StatefinderB}. The cosmographic approach, where the connection between the dynamical history of the Universe and its material components (the Einstein equations) is abandoned, was first suggested in \cite{2004CQGra..21.2603V}, with subsequent work in \cite{2005GReGr..37.1541V,2008PhRvD..78f3501C,2011PhRvD..84l4061C,2015mgm..conf.1574V,Lazkoz:2013by,2008MNRAS.390..210C,2011APh....35...17S,Sollerman09}. In this approach, the expansion history $a(t)$ is reconstructed kinematically in terms of the cosmographic parameters, assuming a Taylor-series expansion of the expansion rate (Hubble parameter today) and its time derivatives. Such a model-independent analysis is more rigorous in testing the statistical significance of the accelerated expansion, and recovering details of the expansion history.

One advantage of a purely kinematic analysis is that the derived quantities will also be similarly model-independent. A good example is the redshift at which the transition from deceleration to acceleration ($z_{\rm acc}$) occurs. There has already been some interest in a fully model-independent determination of this quantity, either through some limited cosmographic analysis \cite{2008MNRAS.390..210C,2011APh....35...17S} or a different presentation of the data through a remapping of the redshift $z$ \cite{2015MNRAS.446.3863S}. Other papers have addressed constraints on the transition redshift when considering models to $\Lambda$CDM \cite{Farooq2013,Farooq2013b,TransitionfR,TransitionfT}. In this paper we use both SN1a and BAO datasets to map the distance-redshift relation expansions in terms of a variable $\zeta$, first introduced in \cite{Cattoen2007}. We then use the statistical constraints on the cosmographic parameters to reconstruct the acceleration history of the Universe as a function of redshift $q(z)$, and determine limits on the redshift of acceleration $z_{acc}$.

While this paper was in preparation, another paper was released claiming strong limits on the redshift of transition \cite{Moresco2016}. However, these constraints exist solely in the context of a $\Lambda$CDM analysis, assuming the Einstein equations. We would argue that this prior (that the $\Lambda$CDM model is the correct model) provides a high degree of constraining power by itself, beyond that of the data. In contrast, a model independent analysis would provide rigorous constraints on the transition redshift.

\section{Cosmographic parameters}

Any cosmographic analysis starts with the assumption of light moving along null geodesics in a homogenous and isotropic universe. The line element is then described by the Friedmann-Lema\^{i}tre-Robertson-Walker metric, which is given by
\be
ds^2=c^2dt^2-a^2(t)\left[\frac{dr^2}{1-\kappa r^2}+d\Omega^2\right] \,,
\ee
where $t$ is cosmological time, $r$ is comoving separation, $\Omega$ is angular separation, and $a(t)$ is the scale factor of the expansion. Here we Taylor expand the scale factor with respect to cosmological time $t$, and define the Hubble $H(t)$, deceleration $q(t)$, jerk $j(t)$, snap, $s(t)$ and lerk $l(t)$, parameters, taking the series expansion to five terms.
\bea
H(t) & = & + \frac1a\frac{\d a}{\d t};  \\
q(t) & = & -\frac1a\frac{\d^2 a}{\d t^2}\left[\frac1a \frac{\d a}{\d t}\right]^{-2};\
\label{eq:qParam} \\
j(t) & = & +\frac1a\frac{\d^3 a}{\d t^3}\left[\frac1a \frac{\d a}{\d t}\right]^{-3}; \\
s(t) & = &  +\frac1a\frac{\d^4 a}{\d t^4}\left[\frac1a \frac{\d a}{\d t}\right]^{-4}; \\
l(t) & = &  +\frac1a\frac{\d^5 a}{\d t^5}\left[\frac1a \frac{\d a}{\d t}\right]^{-5}.
\eea
The Hubble parameter has units of inverse time, while the others are defined to be dimensionless. Using these parameters, we construct the Taylor series of the expansion history
\bea
\frac{a(t)}{a(t_0)} & =&  1+ H_0(t-t_0) - \frac{q_0}{2}H_0^2(t-t_0)^2 + \frac{j_0}{3!}H_0^3(t-t_0)^3  \nonumber\\
&& + \frac{s_0}{4!}H^4_0(t-t_0)^4 + \frac{l_0}{5!}H^5_0(t-t_0)^5 + O[(t-t_0)^6].
\label{eq:scaleFactor}
\eea
The physical distance travelled by a photon that is emitted at time $t$ and absorbed at the current epoch $t_0$ is
\bea
D = c \int dt = c(t_0 - t).
\eea
Hence, the scale factor as a function of the distance is found by substituting the physical distance into equation \ref{eq:scaleFactor},
\bea
\frac{a(t_0 - D/c)}{a(t_0)} & =&  1- \left( \frac{H_0D}{c}\right)  + \frac{q_0}{2}\left( \frac{H_0D}{c}\right) ^2 - \frac{j_0}{3!}\left( \frac{H_0D}{c}\right) ^3   \nonumber \\
& & - \frac{s_0}{4!}\left( \frac{H_0D}{c}\right) ^4 - \frac{l_0}{5!}\left( \frac{H_0D}{c}\right) ^5 \,.
\eea

\subsection{Redshift series expansions}
The cosmographic approach is very effective in directly modelling the distance-redshift relation, however, the fact that it relies on a series expansion means the models have inherent issues. As discussed in \cite{Cattoen2007}, there are two main problems when dealing with series expansions. Firstly, the truncation problem which can be limited by going to more terms in the expansion. However, every additional term brings a new parameter that must be solved. In this paper we consider up to the third, fourth, and fifth derivative of the scale factor (in three separate cases), which appear to map the distance-redshift relation and the deceleration history very effectively. The second problem is the convergence of the series. A redshift larger than zero has an inherent error in the series expansion.

It is possible to limit this convergence problem by parametrising the cosmological distances with a new variable, $\zeta$ \cite{Lazkoz:2013by,Cattoen2007}
\begin{equation}
\zeta = \frac{z}{z+1}.
\end{equation}
This parametrisation translates the z-redshift interval $[0, \infty]$ to a $\zeta$-redshift interval $[0,1]$.
Given $1 - \zeta = \frac{a(t_0 - D/c)}{a(t_0)}$ we can write the distances as a function of $\zeta$,
\bea
\zeta(D) =  \frac{H_0D}{c} - \frac{1}{2}q_0\left(\frac{H_0D}{c}\right)^2 + \frac{1}{6}j_0H_0^3\left(\frac{H_0D}{c}\right)^3 + \frac{1}{24}s_0\left(\frac{H_0D}{c}\right)^4 + \frac{1}{120}l_0\left(\frac{H_0D}{c}\right)^5.
\eea
Then, a reversion of the power series gives:
\bea
D(\zeta) = \frac{c}{H_0} \lbrace \zeta - \frac{1}{2}q_0\zeta^2 +\frac{1}{6}(-j_0+3q_0^2)\zeta^3+\frac{1}{24}(10j_0q_0-15q_0^2+s_0)\zeta^4 \nonumber \\ + \frac{1}{120}(10j_0^2 -l0-105j_0q_0^2+105q_0^4-15q_0s_0)\zeta^5 \rbrace
\label{eq:Dzeta}
\eea

As the SNIa data extends to $z > 1$, the $\zeta$-redshift should limit the convergence problem significantly.

Note that for a given choice of cosmographic parameters the expansion rate could go negative at some time or redshift, leading to a contracting universe. Given the observational evidence against a contracting universe (at least at late times), during our analysis we automatically reject any parameter value that can lead to a contracting universe.

\section{Redshift of transition}
\label{sec:deceleration}

We define the redshift of transition from deceleration to acceleration to be
\be
z=z_{\rm acc} : q(z)=0,\frac{{\rm d}q(z)}{{\rm d}t}<0\,.
\ee
In the case where the Universe is transitioning from deceleration to acceleration more than once, we consider only the latest case (i.e. the lowest redshift where this is true).

The transition redshift is found by determining an expression for the deceleration history of the universe as a function of redshift. Given that the the physical distance travelled by a photon is just the travel time, we can write
\bea
(t-t_0) = -\frac{D}{c}\,.
\eea
Substituting this into the definition of the scale factor and its derivatives  gives the expansion rate and deceleration history of the universe (truncating here at the third term, $j_0$, for ease of presentation):
\bea
a(t) & =& a_0 \left\lbrace  1+ H_0(t-t_0) - \frac{q_0}{2}H_0^2(t-t_0)^2 + \frac{j_0}{6}H_0^3(t-t_0)^3 \right\rbrace .\\
\dot a(t) & =& a_0 \left\lbrace  H_0 - q_0 H_0^2(t-t_0) + \frac{j_0}{2}H_0^3(t-t_0)^2 \right\rbrace. \\
\ddot a(t) & =& a_0 \left\lbrace - q_0 H_0^2 + j_0 H_0^3(t-t_0) \right\rbrace.
\eea
Finally, the deceleration history can be found using the definition:
\bea
q(t) = -\frac{a \ddot a}{\dot a^2}.
\eea
Thus the deceleration history ($q(\zeta)$ or equivalently $q(z)$) can be found by substituting the cosmic time for the distance-redshift relation in the equations for the scale factor and its time derivatives. The transition redshift was found by determining the redshift at which the deceleration crossed from decelerating to accelerating.


\begin{figure}[h!]
\centering
\includegraphics[width=0.7\linewidth]{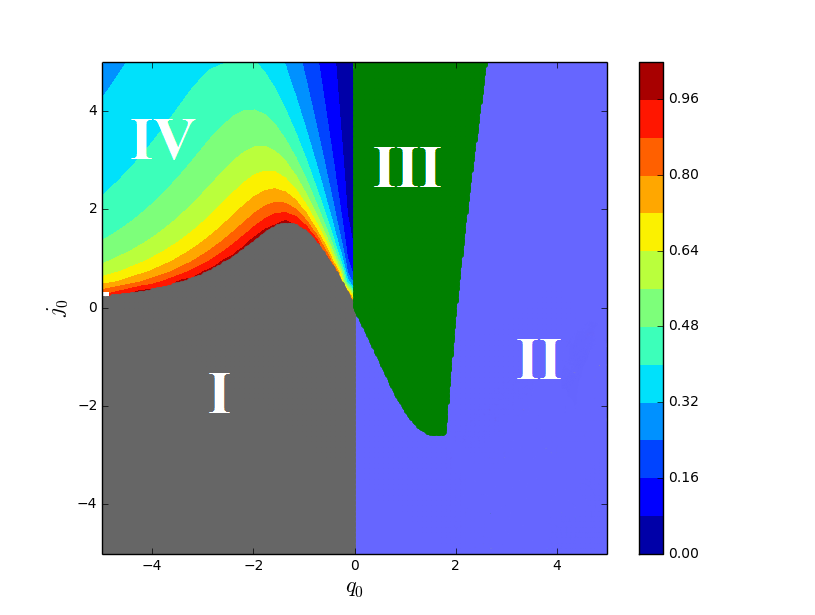}
\caption{Plot of the transition redshift across the $q_0, \ j_0$ parameter space. The plot only considers transitions in the domain $0\le z \le 1$. Region I is accelerating over the whole domain, while Region III is decelerating over the whole domain and does not transition. Region II changes from accelerating at high redshift to decelerating at low redshift. Region IV transitions from decelerating at high redshift to accelerating at a lower redshift in the domain $0\le z \le 1$. The color bar indicates the exact transition redshift for Region IV. }
\label{fig:Comparison}
\end{figure}

 The transition redshift was calculated for a range of parameters in the $q_0$ and $j_0$ parameter space only considering transitions in the domain $0\le z \le 1$. A plot of the transition redshift comparison figure is illustrated in Figure \ref{fig:Comparison}. The plot is separated into four distinct regions of transition.

\begin{description}
	\item{Region I:} \hfill \\
		This region does not transition, and is accelerating at least over whole domain $0\le z \le 1$.
	\item{Region II:} \hfill \\
		This region was accelerating at redshift $z = 1$ and transitioned to decelerating at a lower redshift.
	\item{Region III:} \hfill \\
		This region does not transition and is decelerating at least over the whole domain $0\le z \le 1$.
	\item{Region IV:} \hfill \\
		This region transitions from decelerating at high redshift to accelerating at a lower redshift. The transition redshift is indicated by the colour bar at right.
\end{description}

\begin{figure}[h!]
		\centering
		\includegraphics[width=0.8\linewidth]{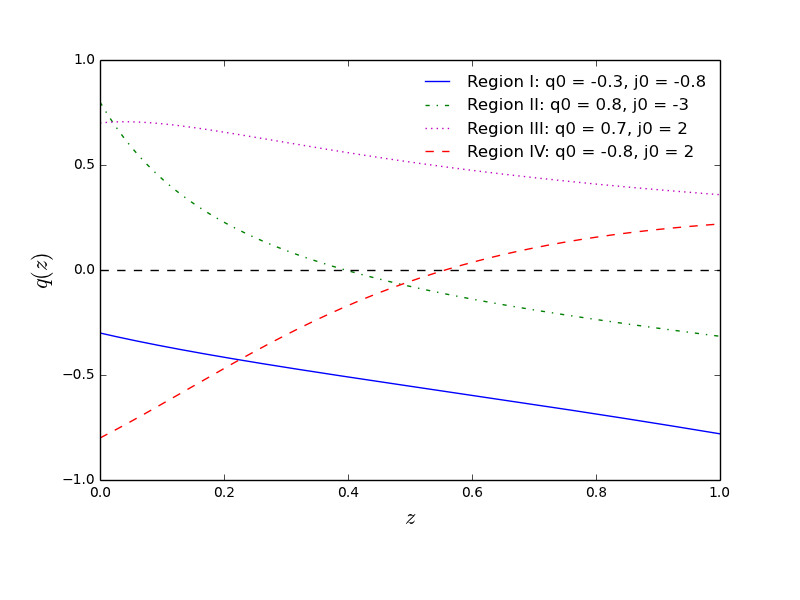}
\caption{Example plots of the deceleration history of the universe $q(z)$ with parameters  illustrated on the figures. Each line depicts an example of the behaviour from a region of the parameter space in Figure \ref{fig:Comparison}. Notice that only region IV corresponds to the kind of behaviour we would expect in a $\Lambda$CDM universe. If we assume a prior of deceleration by $z=1$, then regions I \& II would automatically be ruled out.}
\label{fig:TransitionRedshiftExample}
\end{figure}

As an example, Figure \ref{fig:TransitionRedshiftExample} shows the deceleration history against redshift for 4 different sets of parameter pairs ($q_0,~j_0$), with $s_0$ and $l_0$ set to zero.

\subsection{Prior assumption of early deceleration}
If we make the prior assumption that the universe was decelerating at a high redshift (as would be required, for example, to fit the data from the Cosmic Microwave Background), we might require that at redshift $z = 1.0$ the deceleration must be $q_{(z=1.0)} > 0$. This significantly limits the parameter space available for the analysis. For example, in the $q_0,~j_0$ parameter space shown in figure \ref{fig:Comparison}, we would  remove regions I and II as possibilities, as they both are accelerating at high redshifts. Regions III and IV would remain in the available parameter space, however, because they both were decelerating at high redshifts. The difference is that region III has always been decelerating, and represents a currently decelerating universe, whereas region IV represents a currently accelerating universe that transitioned between $0\le z \le 1$. For a higher-dimensional parameter space, where $s_0$ or $s_0$ and $l_0$, were also allowed to vary, the available parameter would also be smaller, by the same argument.

\section{Analysis method}

\subsection{Supernovae Type-Ia data}
Following the method outlined in \cite{2011PhRvD..84l4061C}, the $H_0$-independent luminosity distance expanded in $\zeta$ is given in equation \ref{eq:zetadL}. We note an error in the second term of the authors' expansion, and have fixed it in our expression. In the cases where we are only extending our series to the jerk parameter, $j_0$, we still expand both series' to five terms due to the strong bias caused by using only three terms. Since we are only approximating an inversion of a power series, there are no new parameters introduced by increasing the order of our series.

The distance modulus is then calculated for each of the series expansions using equation \ref{eq:mu}
\bea
\mu = 5\log_{10} \left( {d_L}' \right) .
\label{eq:mu}
\eea
Where $d_L'$ is the $H_0$-independent luminosity distance calculated from the series expansions. In our analysis we make use of the latest Union2.1 compilation of SNIa \cite{2012ApJ...746...85S}. The Union2.1 sample includes 580 SNIa events distributed over the redshift interval $0.015 \le z \le 1.414$. In order to compare the models against the data, we calculate the $\chi^2$ function following the method outlined in \cite{Goliath:2001} where we marginalise over the Hubble constant, $H_0$, and consider the covariance of the correlated data.

\subsection{Baryonic Acoustic Oscillation data}
We applied a similar cosmographic approach to derive series expansions for $D_V$, where $D_V$ is a composite of the physical angular diameter distance and Hubble parameter. Using equation \ref{eq:Dv}, the series expansion in $\zeta$ is given in equations \ref{eq:zetaDV}.
\bea
D_V = \left[ \frac{d_L^2}{(1+z)^2} \frac{cz}{H(z)} \right]^{1/3}
\label{eq:Dv}
\eea
In the same way as for the luminosity distance, we extend the series to five terms. This does not introduce any new parameters as we are still using only up to the third derivative of the scale factor in the series expansion.

In our analysis we use the latest BAO distance dataset including the 6dFGS \cite{beutler:2011}, SDSS\cite{SDSSMGS}, BOSS \cite{anderson2014} and WiggleZ\cite{kazin2014} surveys which together comprise measurements at seven different redshifts in the range $0.106 \le z \le 0.73$. A summary of the measurements of the distilled parameters $d_z$ and $f(z)$ are quoted in Table \ref{tab:BAOdataset}.
\begin{table}[h]
\centering
\begin{tabular}{|c|c|c|c|}
\hline Sample	& z 	& $D_V/r_d$ 				& $f(z)$ \\
\hline 6dFGS	& 0.106 & $3.05 \pm 0.137$ 	&  \\
 SDSS-MGS 	& 0.15	& $4.48 \pm 0.17$ 	& $1.47 \pm 0.08$ \\
 BOSS-LOWZ 	& 0.32 	& $8.47 \pm 0.17$ 	& $2.78 \pm 0.13$ \\
 BOSS-CMASS 	& 0.57 	& $13.77 \pm 0.14$ 	& $4.52\pm 0.21$ \\
 WiggleZ w/recon 	& 0.44 	& $11.50 \pm 0.55$ 	& $3.77 \pm 0.25$ \\
 WiggleZ w/recon	& 0.6 	& $14.88 \pm 0.67$ 	& $4.88 \pm 0.31$ \\
 WiggleZ w/recon	& 0.73 	& $16.86 \pm 0.57$ 	& $5.53 \pm 0.31$ \\
\hline
\end{tabular}
\caption{\label{tab:BAOdataset} BAO distance dataset from the 6dFGS, SDSS, and BOSS as summarised in \cite{Aubourg2015}, and WiggleZ Survey as given in \cite{kazin2014}. Measurements of the parameter $d_z$ and the calculated value of $f$ for each of the samples at the corresponding redshift, $z$, is shown in the table. $f$ was calculated using equation \ref{eq:f}. The correlation of the $f$ parameter is illustrated in Table \ref{tab:fcov}.}
\end{table}

We define the $D_V$ parameter as being related to the measured $d_z$ values by the following equation,
\begin{equation}
D_V = \frac{r_s}{d_z}
\end{equation}
where $r_s$ is the baryonic acoustic distance. In order to eliminate this parameter from our analysis, we instead determine a ratio of $D_V$ over the 6dFGS $D_V$ parameter, $f$, defined as,
\begin{equation}
f = \frac{D_V}{D_V^{\rm 6dFGS}}
\end{equation}
The benefit of using this relationship is that the fraction simplifies to a ratio of the measured $d_z$ parameters as shown in the following equation.
\begin{equation}
f = \frac{{d_z^{\rm6dFGS}}}{d_z}
\label{eq:f}
\end{equation}

\begin{figure}[h]
\centering
\includegraphics[width=0.7\linewidth]{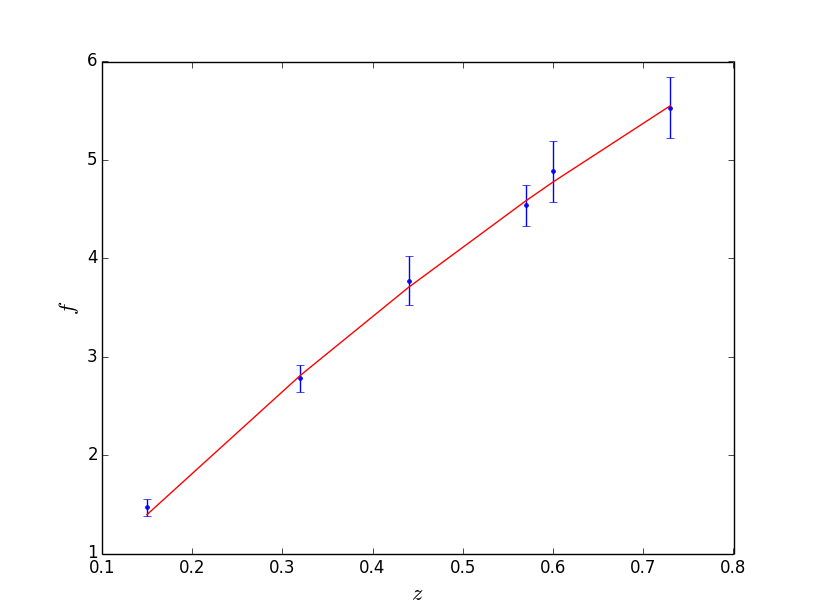}
\caption{Plot of the normalised $f=D_V/D_{V}^{{\rm 6dfGS}}$ parameter of BAO against the redshift. The red trend line is modelled with cosmographic parameters $q_0 = -0.56$, $j_0 = 1$.}
\label{fig:f_vs_z}
\end{figure}

A plot of $f$ against the redshift with a cosmographic model superposed is illustrated in Figure \ref{fig:f_vs_z}.

The covariance matrix of the $f$ parameter was constructed by using the correlation coefficients for each of the parameter sets given in \cite{Percival:2010,blake:2011dm}.
The off diagonal covariance matrix entries were calculated using 
\begin{equation}
\label{eq:cov_entry}
C_{x y} = r_{x y}\sigma_x \sigma_y \,,
\end{equation}
where $r_{x y}$ are the correlation coefficients of the particular datasets 
and $\sigma_{x}$, $\sigma_{y}$  are the uncertainties of  the $f$ parameters $x$ and $y$ listed in Table \ref{tab:BAOdataset}. The diagonal entries were calculated as simply the square of the uncertainties of each $f$ parameter listed in Table \ref{tab:BAOdataset}.


\begin{table}[h!]
\centering
\begin{tabular}{|c|c c c c c c|}
\hline Redshift & $z_{0.15}$  & $z_{0.32}$ & $z_{0.57}$ & $z_{0.44}$ & $z_{0.6}$ & $z_{0.73}$ \\
\hline $z_{0.15}$ & 134.95 & 0 & 0 & 0 & 0 & 0 \\
 $z_{0.32}$  & 0 & 53.72 & 0 & 0 & 0 & 0  \\
 $z_{0.57}$ & 0 & 0 & 23.12 & 0 & 0 & 0  \\
 $z_{0.44}$  & 0& 0 & 0 & 24.33 & -12.01 & 4.29 \\
 $z_{0.6}$ & 0 & 0& 0 & -12.01 & 17.83 & -6.37 \\
 $z_{0.73}$  & 0& 0 &  0 & 4.29 & -6.37 & 12.56 \\
\hline
\end{tabular}
\caption{\label{tab:fcov} The combined inverse covariance matrix $C^{-1} $ of the $f$ parameter measurements. 
}
\end{table}

In order to compare the BAO data against our models, we calculate the ${\chi}^2$ function with the following equation
\begin{equation}
{\chi}^2 = (\Delta^T) C^{-1} (\Delta)
\end{equation}
where $\Delta$ is a column matrix representing the difference between the data and the model's prediction. We again use the Metropolis Hastings Markov Chain Monte Carlo (MCMC) method to map the statistical confidence intervals and hence determine the best fit parameters.

\subsection{Models and parameters priors}
\label{sec:cases}
We consider three cases for our analysis:
\begin{enumerate}[i]
	\item Only the deceleration $q_0$ and the the jerk $j_0$ parameters are allowed to vary, and all other cosmographic parameters ($s_0$ and $l_0$) are fixed at zero.
	\item The deceleration $q_0$, the jerk $j_0$, and the snap $s_0$ are allowed to vary, and the lerk $l_0$ is fixed at zero.
	\item All four cosmographic parameters considered in this analysis ($q_0,~j_0,~s_0,~l_0$) are allowed to vary.
\end{enumerate}
We then use a Metropolis Hastings MCMC method to map the statistical confidence intervals over the available parameter space. We use MCMC Hammer\footnote{\href{http://dan.iel.fm/emcee/}{dan.iel.fm/emcee}.} \cite{ForemanMackey:2012}, a Python implementation of the affine-invariant ensemble sampler proposed by \cite{GoodmanWeare}. We initially consider no priors on the parameter ranges, and allow them to explore the entire available parameter range.  However, for each case we will also subsample the chain in post-processing, to impose the prior of deceleration before $z=1$, and investigate how this changes the posterior probability distribution.

\section{Results}

\subsection{Cosmographic parameters}
\label{sec:resultsparameters}
\begin{figure}[h]
\centering
\includegraphics[width=.6\textwidth]{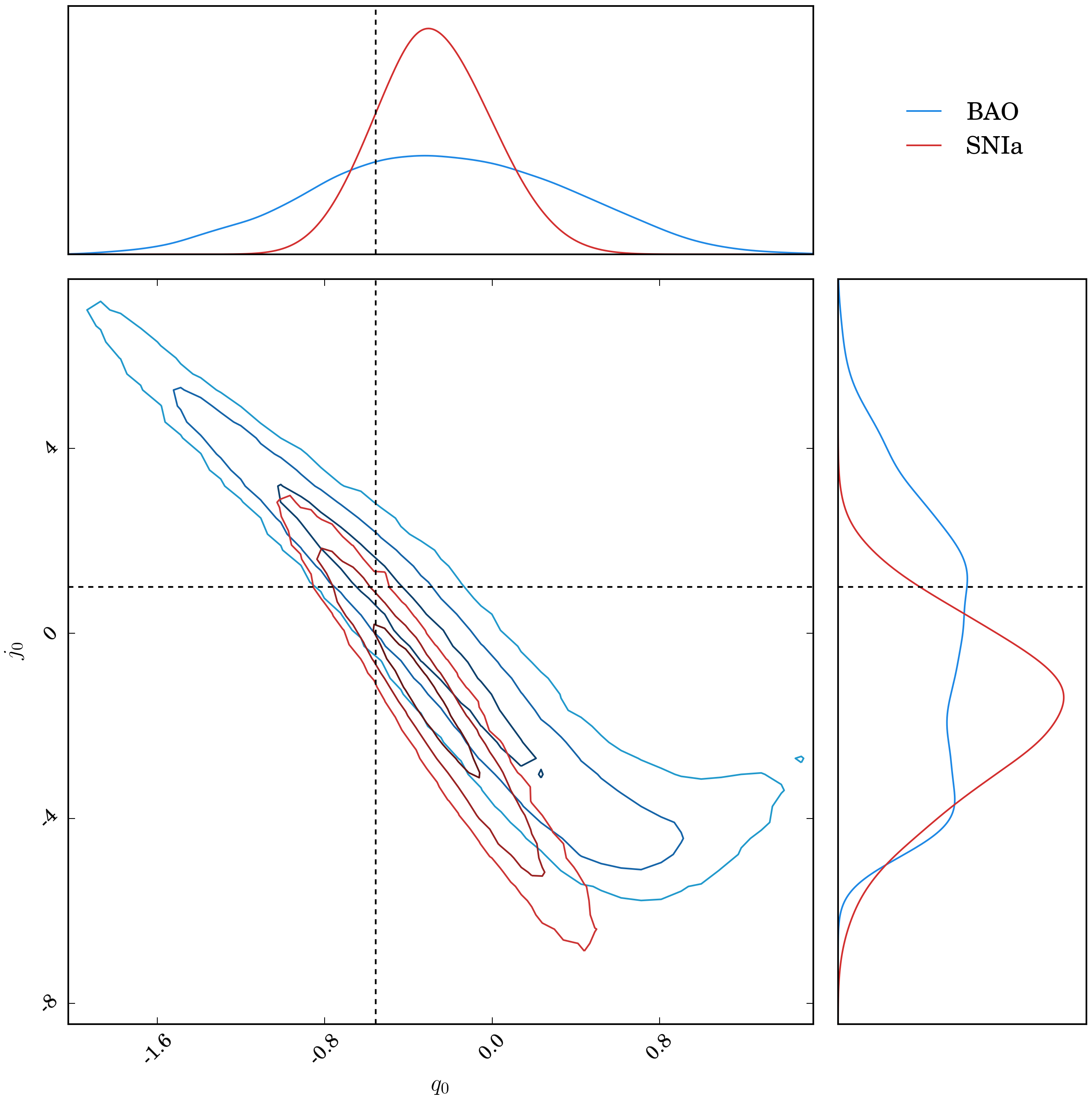}
\caption{Confidence intervals of the deceleration $(q_0)$ and jerk $(j_0)$ cosmographic parameters for the SNIa (red) and BAO (blue) data. The contour lines represent $68.3\%$, $95\%$ and $99.{\tiny }7\%$ confidence regions respectively.
The lines mark the positions of the $\Lambda$CDM model prediction $(q_0=-0.56, j_0=1)$.}
\label{fig:SNIaandBAO_zeta}
\end{figure}
\begin{figure}[h]
\centering
\includegraphics[width=.8\textwidth]{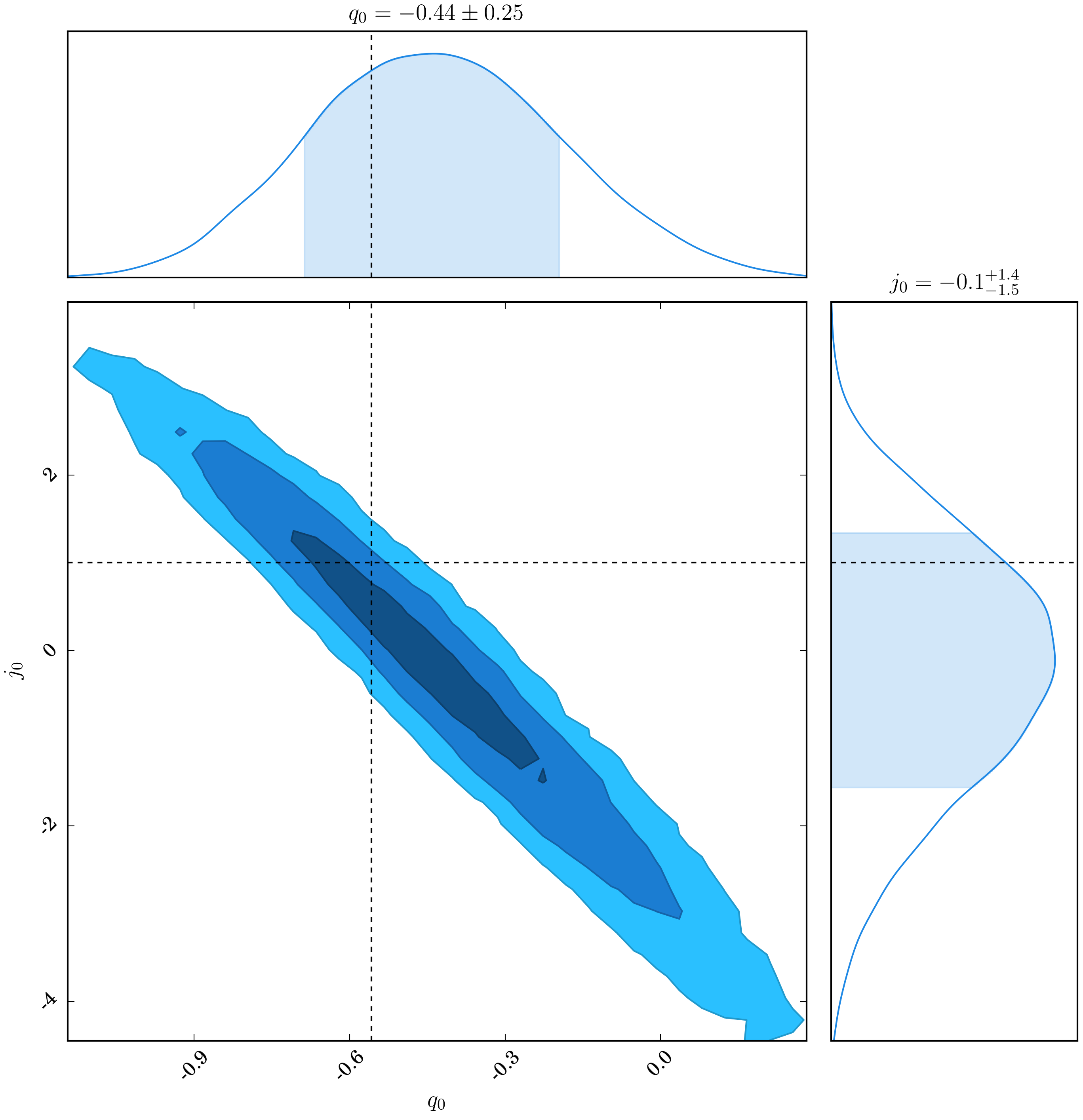}
\caption{Probability contours of the deceleration and jerk parameters from combining the SNIa and BAO data sets. The contour lines represent $68.3\%$, $95\%$ and $99.{\tiny }7\%$ confidence regions respectively.
The lines mark the positions of the $\Lambda$CDM model prediction $(q_0=-0.56, j_0=1)$.}
\label{fig:SNIaandBAO_zeta_combined}
\end{figure}
The marginalised parameter constraints for different data combinations and cosmographic expansions are listed in table \ref{tab:Combined}. In the cases where the constraints on $s_0$ and $l_0$ are not listed, these parameters were not varied, but set to zero.

\begin{table}[h]
\centering
\begin{tabular}{|l|c|c|c|c|c|}
\hline
              &Case    &  $q_0$   & $j_0$ & $s_0$ & $l_0$ \\ \hline
 \multirow{3}{*}{SN-Ia}   & i  & $-0.30^{+0.28}_{-0.26}$ & $-1.2^{+1.5}_{-2.1}$ & -- & -- \\
      & ii& $0.01^{+0.21}_{-0.23}$ & $-14.3^{+13.3}_{-6.6}$ & $9^{+64}_{-83}$  &-- \\
       & iii & $-0.63^{+0.83}_{-0.52}$ & $0.6^{+14.7}_{-18.6}$ & $-11.9^{+199.0}_{-383.6}$ & $ 700^{+5800}_{-4900} $  \\ \hline
  \multirow{3}{*}{BAO}    & i & $-0.32^{+0.70}_{-0.53}$ & $1.16^{+0.78}_{-5.13}$    & -- & -- \\
        & ii& $-0.21^{+0.64}_{-0.57}$ & $-2.7^{+10.3}_{-10.5}$ & $45.4^{+45.9}_{-214.6}$& -- \\
       & iii & $-0.28^{+1.33}_{-0.89}$ & $-3.7^{+20.1}_{-17.9}$ & $71.2^{+275.3}_{-220.9}$ & $\ 400^{+3500}_{-1700} $  \\ \hline
 \multirow{3}{*}{Combined}& i & $-0.41^{+0.21}_{-0.28}$ & $-0.1\pm 1.4$ & -- & -- \\
 & ii& $-0.24^{+0.15}_{-0.29}$ & $-2.7^{+3.1}_{-4.3}$ & $-16.4^{+33.0}_{-43.5}$ & --\\
 & iii & $-0.25^{+0.49}_{-0.43}$ & $-1.6^{+8.7}_{-8.3}$ & $29.1^{+106.1}_{-77.4}$ & $860^{+1108}_{-896}$   \\  \hline
\end{tabular}
\caption{\label{tab:Combined}The mean and 1-sigma error bars for the cosmographic parameters for each of the data sets and parameter combination cases defined in section \ref{sec:cases}. }
\end{table}

\begin{figure}[h]
\centering
\includegraphics[width=.8\textwidth]{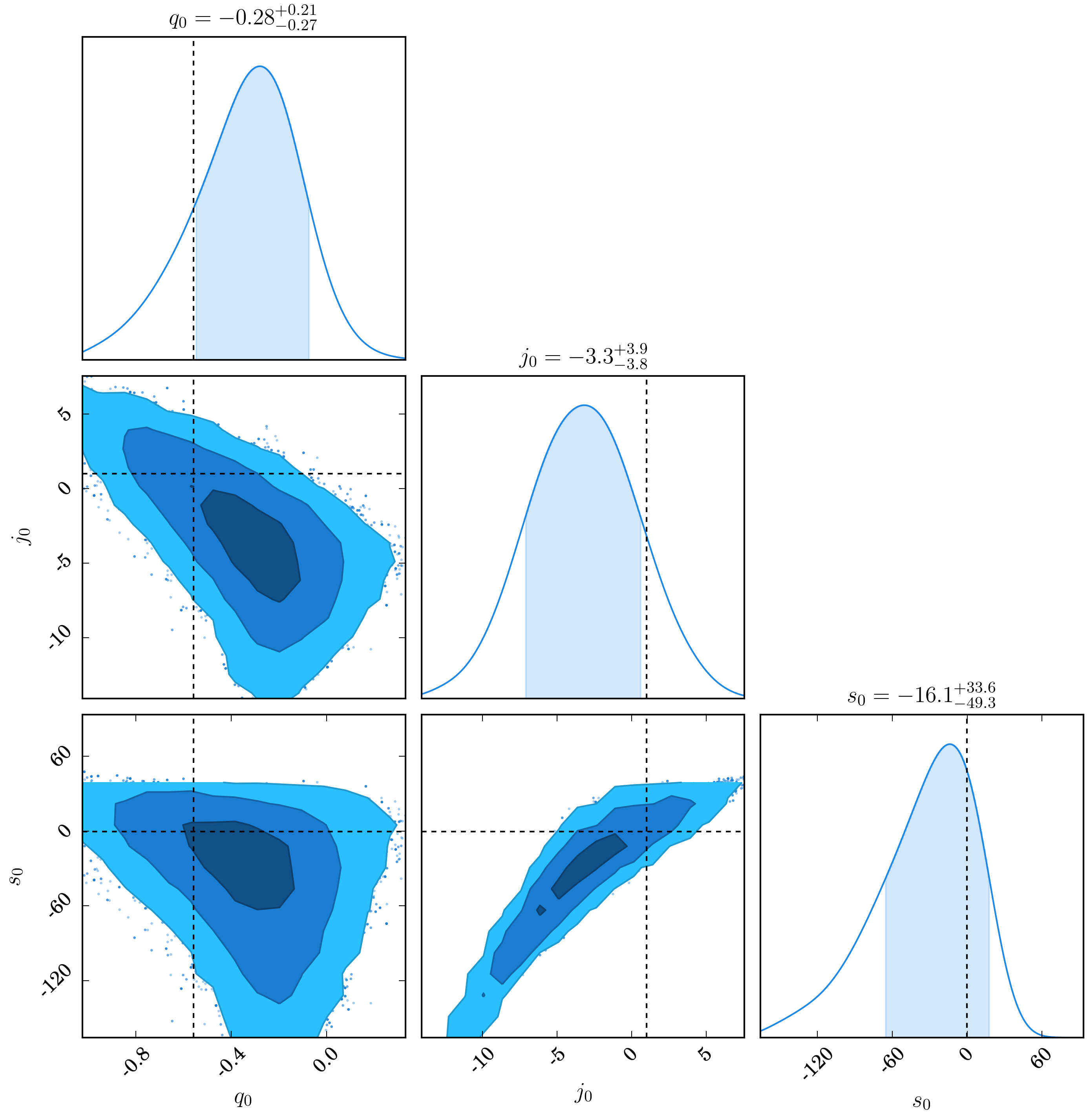}
\caption{MCMC chain elements and probability contours of $(q_0)$, jerk $(j_0)$ and snap ($s_0$)  parameters using the combined data sets. The contour lines represent $68.3\%$, $95\%$ and $99.7\%$ confidence regions respectively. The vertical/horizontal lines denote  the values of the cosmographic parameters as predicted by $\Lambda$CDM. }
\label{fig:combined_zeta_q0j0s0}
\end{figure}

\begin{figure}[h]
\centering
\includegraphics[width=.8\textwidth]{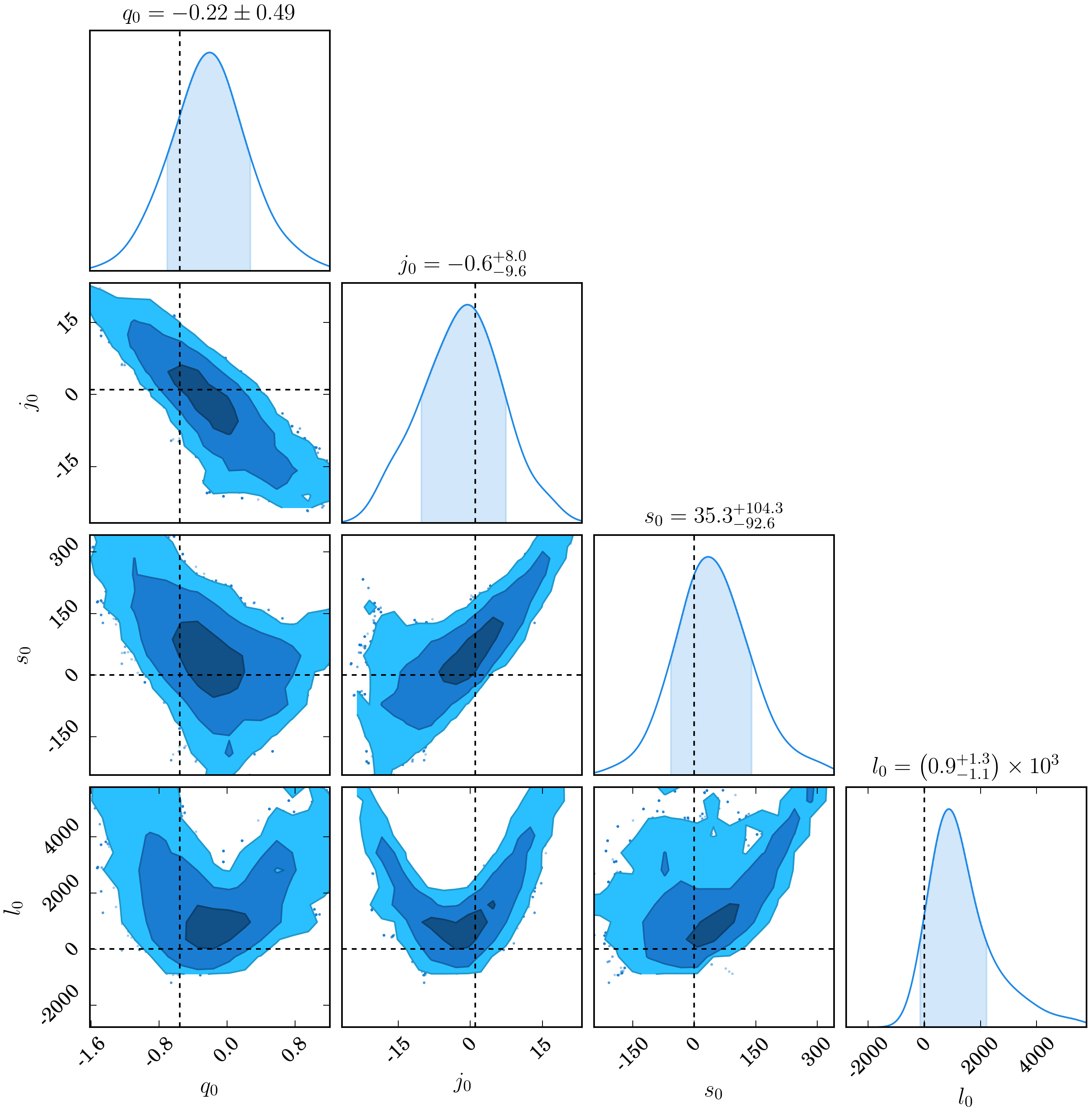}
\caption{MCMC chain elements and probability contours of the  deceleration $(q_0)$, jerk $(j_0)$, snap ($s_0$) and ($l_0$) for the combined SNIa and BAO and data.  The contour lines represent $68.3\%$, $95\%$ and $99.7\%$ confidence regions respectively. The vertical/horizontal lines denote  the values of the cosmographic parameters as predicted by $\Lambda$CDM. }
\label{fig:combined_zeta_q0j0s0l0}
\end{figure}

The two-dimensional probability distributions for the $q_0$ and $j_0$  analysis is illustrated in Figure \ref{fig:SNIaandBAO_zeta}, where we show the individual contributions from the two different datasets, and Figure \ref{fig:SNIaandBAO_zeta_combined}, where we show the final contour when combining the two datasets.  The one and two-dimensional marginalised probability distributions for the $q_0$, $j_0$ and $s_0$  analysis are illustrated in Figure \ref{fig:combined_zeta_q0j0s0}. Finally the one and two-dimensional marginalised probability distributions for the $q_0$, $j_0$, $s_0$ and $l_0$  analysis are illustrated in Figure \ref{fig:combined_zeta_q0j0s0l0}.  In all cases we see that the value of the cosmographic parameters predicted in the $\Lambda$CDM model lie within the 95\% probability contour.


Firstly we see  that the supernovae data does a lot better at constraining the cosmographic parameters than the BAO data. In all cases the one-sigma errors listed in table \ref{tab:Combined} are equal to or larger when using the BAO data alone compared to using the SN-Ia data alone. The combination of SN-Ia and BAO data is always an improvement compared to the individual data alone, so adding the BAO data to the supernovae data does improve on the constraints.

Secondly we see that the constraints on the cosmographic parameters are sensitive to the order of the expansion in $a(t)$. The deceleration  parameter today $q_0$ is the least sensitive of these. For example, we see from table \ref{tab:Combined} that when only two parameters are varied in the combined dataset the limits on the deceleration parameter are $q_0=-0.41^{+0.21}_{-0.28}$, however, when we increase to four parameters these errors double to $q_0= -0.16^{+0.41}_{-0.56}$. In contrast, constraints on the jerk change from $j_0=-0.1\pm1.4$ (when only two parameters are varied) to  $j_0 = 2.2^{+5.7}_{-12.6}$ (when four parameters are varied), with the errors increasing by a factor of five or more. As you can see from figures \ref{fig:combined_zeta_q0j0s0} and \ref{fig:combined_zeta_q0j0s0l0}, the large range of possible values for the jerk are driven by degeneracies with the snap $s_0$ and lerk $l_0$ parameters, which are also allowed by the data to take extreme values. These extreme values for $s_0$ and $l_0$ may not be physical, and will most probably be ruled out by more data.

\subsection{Deceleration history}

\begin{figure}[h]
\centering
\includegraphics[width=.45\textwidth]{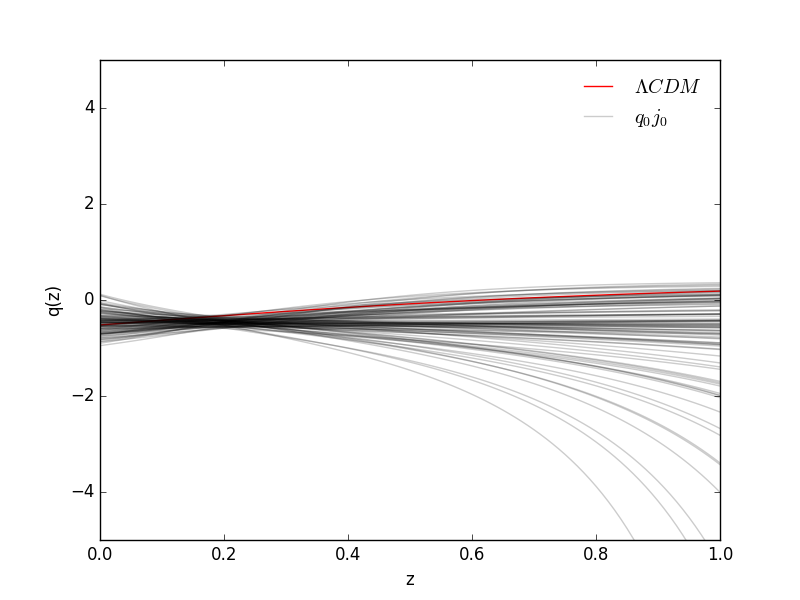}\\
\includegraphics[width=.45\textwidth]{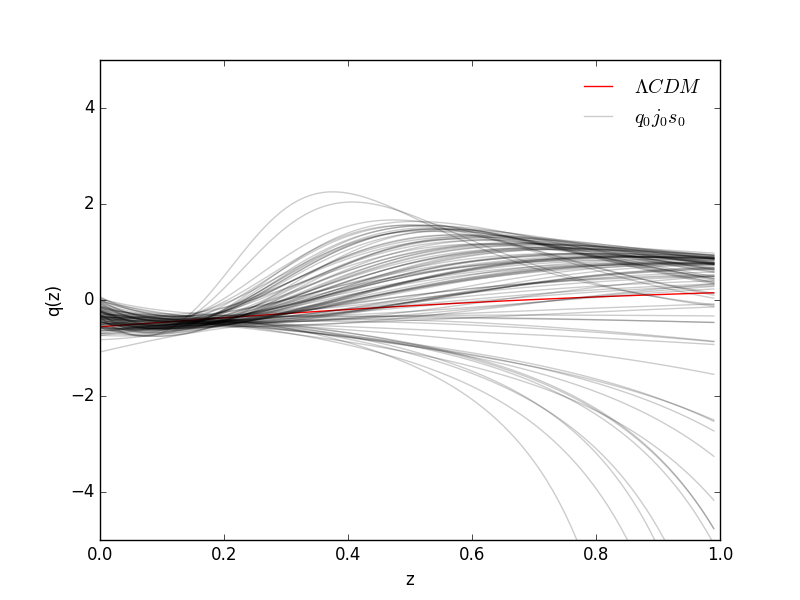}\\
\includegraphics[width=.45\textwidth]{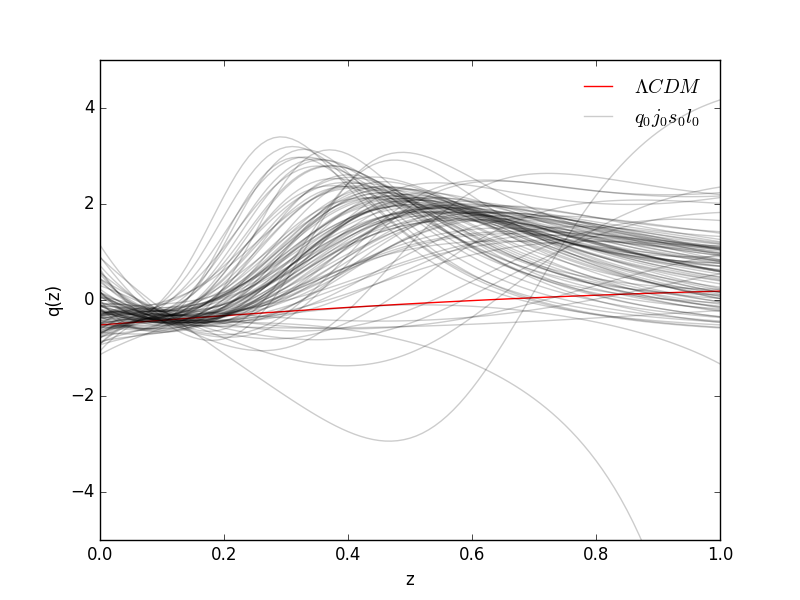}
\caption{Deceleration histories as a function of redshift $z$ for a representative sample of 100 chain elements}
\label{fig:decelerationhistory}
\end{figure}

We can place statistical limits on the form of the deceleration history by sampling different regions of the parameter space by their posterior probability. In figure \ref{fig:decelerationhistory} we show 100 random $q(z)$ functions as predicted by their parameter values taken from the MCMC chain.

For the first case, where only $q_0$ and $j_0$ are allowed to vary, we see a fairly smooth history, with no sudden jumps or transitions.  As we further expand the taylor series for $a(t)$, $\dot{a}(t)$ and $\ddot{a}(t)$ by adding in the snap $s_0$ and lerk $l_0$ parameters, we allow the deceleration history to take on a more complex form, with many curves that rise quickly in $q(z)$ before reaching a maximum and decreasing. Even so, in all of these cases the deceleration parameter is well constrained in the region before $z \sim 0.1$, but becomes much less constrained as z increases towards $z=1$. Indeed in many such cases, the deceleration has been negative in the past, but has transitioned back to being positive at very late times, before becoming negative once again. In these cases we only consider the transition from positive to negative, i.e. the onset of the latest period of acceleration, as being the redshift of transition.

\subsubsection{Early deceleration prior}

We now consider the effect of adding a prior, such that the Universe has to be decelerating by $z=1$. We do this by importance sampling the MCMC chains from the combined analysis reported in section \ref{sec:resultsparameters}, assigning a probability of zero to all chain elements that violate this prior. The effect on the marginalised one-dimensional parameter constraints is listed in table \ref{tab:Combinedprior}.

\begin{table}[h]
\centering
\begin{tabular}{|l|c|c|c|c|c|}
\hline
                  & Case &  $q_0$   & $j_0$ & $s_0$ & $l_0$ \\ \hline
 \multirow{3}{*}{Combined -  with prior} & i & $ -0.75^{+0.10}_{-0.12}$ & $1.66^{+0.59}_{-0.37}$ & -- & -- \\
 & ii & $-0.30^{+0.30}_{-0.45}$ & $-4.45^{+5.17}_{-6.41}$ & $-43^{+38}_{-110}$ & --\\
 & iii & $-0.36^{+0.30}_{-0.41}$ & $1.51^{+6.88}_{-8.53}$ & $70.1^{+93.2}_{-91.6}$ & $1188^{+1496}_{-711}$  \\  \hline
\end{tabular}
\caption{\label{tab:Combinedprior}The mean and 1-sigma error bars for the cosmographic parameters for each of the data sets and parameter combination cases defined in \ref{sec:cases}, assuming a prior that the Universe is decelerating by $z=1$. }
\end{table}

We find that the effect of such a prior gives a pronounced difference when only the deceleration $q_0$, jerk $j_0$ and snap $s_0$ are varied. For example, comparing tables \ref{tab:Combined} and \ref{tab:Combinedprior} in the two parameter case, the error bars on $q_0$ decrease by around half, and for the jerk, the errors are a third of those without the prior. However, when the expansion is increased up to the lerk $l_0$, the one dimensional parameter constraints are very similar to those without the prior. This indicates that in the four parameter case there is enough flexibility for almost all the models to be decelerating, and such a prior has limited effect.

\subsection{Transition constraints}

\begin{figure}[h]
\centering
\includegraphics[width=.45\textwidth]{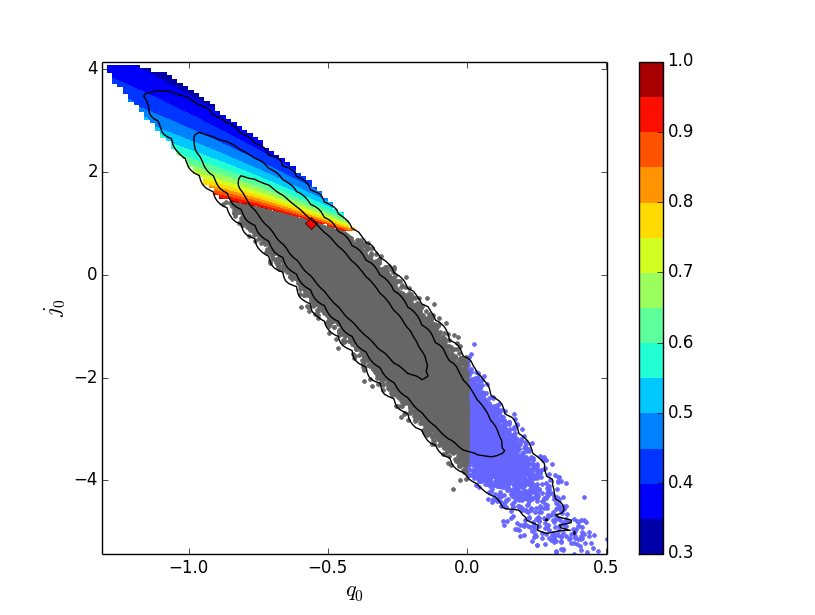}
\caption{The two dimensional posterior probability contours for the case where only the deceleration $q_0$ and jerk $j_0$ parameters are allowed to vary plotted over the transition redshift from Fig. \ref{fig:Comparison}. Much of the posterior parameter space is taken up by `always accelerating' (grey, in the centre) or `decelerating-to-accelerating' (purple, in the bottom right) models, with only a small region in the top-left with `acclerating-to-decelerating' models. The colour bar shows the redshift of transition $z_{acc}$ for these models.}
\label{fig:TransitionConstraints}
\end{figure}

The redshift of transition can be plotted with the confidence contours in the $q_0$ and $j_0$ parameter space in Figure \ref{fig:TransitionConstraints} for case (i) where only $q_0$ and $j_0$ are allowed to vary. These plots give an indication of the best fit transition redshift. The figure shows that much of the posterior parameter space is taken up by models which are either still accelerating at $z=1$ (described as Region I in section \ref{sec:deceleration}) or else were accelerating at high redshift and transitioned in the opposite direction to be decelerating today (Region II). However, we noticed that for models transitioning from deceleration at high redshift to acceleration at low redshift, a very low redshift of transition $z_{acc}<0.3$ is ruled out at 3-sigma.

\begin{figure}[h]
\centering
\includegraphics[width=.45\textwidth]{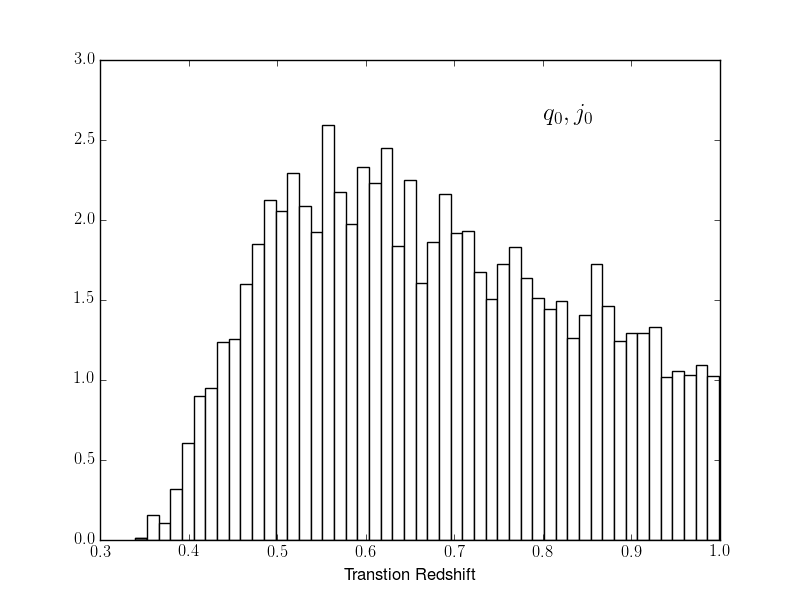}\\
\includegraphics[width=.45\textwidth]{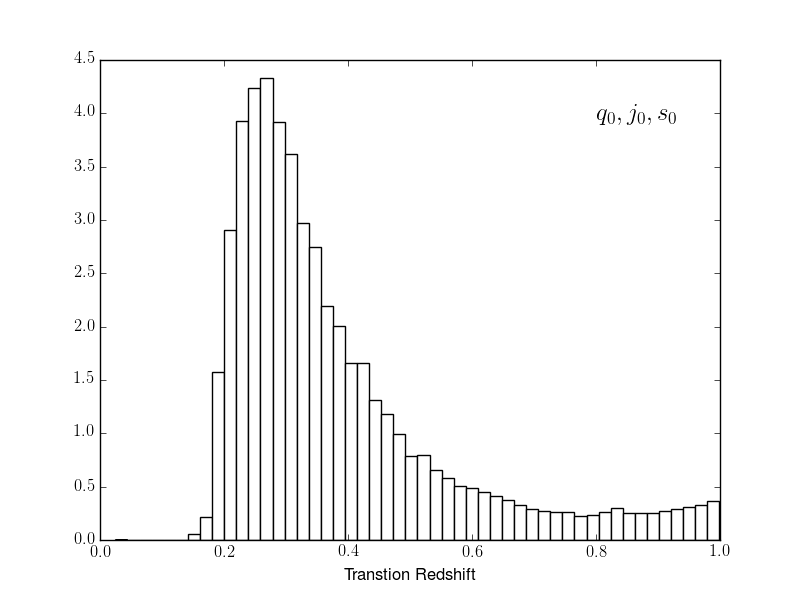}\\
\includegraphics[width=.45\textwidth]{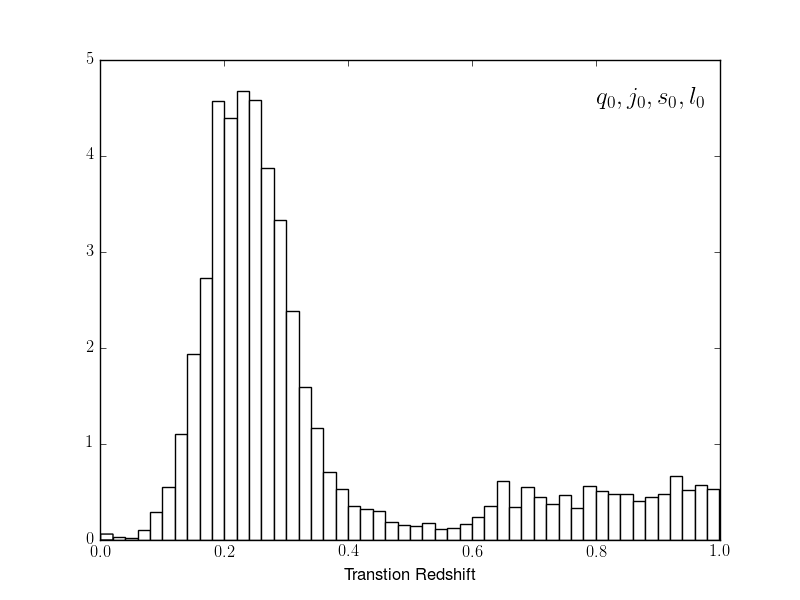}
\caption{Probability distribution of the redshift of transition to acceleration, for different series expansions of $a(t)$. We see that the distribution changes as we go to higher-order expansions, with the analysis that includes the snap ($s_0$) and lerk ($l_0$) parameters preferring an abrupt transition to acceleration at low redshift.
\label{fig:TransitionConstraintsHistograms}}
\end{figure}

For higher dimensional models, with more than two parameters, it is difficult to overplot the transition redshift on the marginalised parameter constraints, as the degeneracies wash out any details such a figure would contain. Instead, in Figure \ref{fig:TransitionConstraintsHistograms} we plot the one-dimensional probability distribution for the redshift of acceleration $z_{\rm acc}$ for the three different analysis cases.  In table \ref{tab:zacclimits} we give the 95\% confidence lower limit on the redshift of transition.

\begin{table}[h]
\centering
\begin{tabular}{|l|c|}
\hline
Model & Transition limit \\ \hline
$q_0, j_0$ & $z_{\rm acc} > 0.44$ \\
$q_0, j_0, s_0$ & $z_{\rm acc} > 0.21$ \\
$q_0, j_0, s_0, l_0$ & $z_{\rm acc} > 0.14$ \\ \hline
\end{tabular}
\caption{\label{tab:zacclimits} Lower limits on the redshift of acceleration, at 95\% confidence.}
\end{table}
The distribution and lower limit on $z_{\rm acc}$ change for different expansion series, seemingly converging on $z=0$ as the series is increased. However, the region of parameter space that correspond to these abrupt low-redshift transitions are models with very large values of the snap and lerk parameters. If a prior is placed on the snap and lerk, or these parameters are better constrained by future data, the constraints become the same as for the lower dimensional models.

\section{Summary}

We  have undertaken a cosmographic analysis of standard candle and standard ruler data in order to determine a model-independent estimate of the epoch of the onset of acceleration, as given by the redshift $z_{acc}$. We demonstrate that a large volume of the available parameter space is inconsistent with a decelerating universe at high-redshift. Even the modest prior of deceleration by $z=1$ severely limits the available parameter space.

In our likelihood analysis, we have used type-Ia supernovae data from the Union 2.1 data compilation  \cite{2012ApJ...746...85S} as our standard candle data set, and Baryon Acoustic Oscillation data from the 6dFGS, \cite{beutler:2011}, WiggleZ Dark Energy Survey \cite{blake:2011dm} and SDSS \cite{Percival:2010} surveys as our standard ruler data set. For the BAO dataset we standardize our rulers relative to the lowest redshift measurement (6dfGS), rather than the sound horizon at high redshift (as measured by the CMB), so as to remove any assumptions regarding the details of the expansion history at high-redshift. We perform a Markov Chain Monte Carlo analysis, marginalising over the value of $H_0$ when using the SN-Ia data.

We performed three different analyses, increasing the number of terms in the expansion of the scale factor $a(t)$ in each case. In the first case with only two cosmographic parameters, we find that $q_0$ and $j_0$ are well constrained by a combination of the BAO and SN-Ia data. As we increase to three parameters $q_0$, $j_0$ and $s_0$, we find that the constraints on $q_0$ remain roughly the same, while the constraints on $j_0$ widen substantially. This is is also the case when we increase to four parameters $q_0$, $j_0$, $s_0$ and $l_0$. All our constraints are consistent with the values of the cosmographic parameters as predicted by the $\Lambda$CDM cosmological model at the 95\% confidence level.

By sampling from our MCMC chains, we show the that acceleration is well constrained at very low redshift ($z<0.1$), independent to the order of the expansion. However, for $z>0.1$, the statistical limits on the deceleration $q(z)$ are very model dependent, and the data allows for a very abrupt transition to acceleration at low-redshift for the higher order models. This rapid transition models are driven by very large values of the snap $s_0$ and lerk $l_0$ parameters, that may be ruled out by future data.

\section*{Acknowledgements}
DP is supported by an Australian Research Council Future Fellowship [grant number FT130101086]. The figures in this paper were generated using the ChainConsumer\footnote{\href{https://samreay.github.io/ChainConsumer/}{samreay.github.io/ChainConsumer/}.} package developed by Samuel Hinton. We thank Tamara Davis and Chris Blake for helpful comments on the draft.

\appendix

\section{Appendix A: SNIa Luminosity distance series expansion}

Using the definition $\zeta = \frac{z}{z+1}$ we obtain a series expansion in zeta. We note an error in \cite{2011PhRvD..84l4061C} in the second term of its expansion, and have fixed it here. We extend our expansion in zeta to five terms due to the bias caused by using only three terms. Since we are using the scale factor up to only the second derivative in our analysis, the number of terms in the expansion does not introduce any new parameters.
\begin{equation}
\label{eq:zetadL}
{d_L}' = {d_1}_\zeta \zeta + {d_2}_\zeta \zeta^2 + {d_3}_\zeta \zeta^3 + {d_4}_\zeta \zeta^4 + {d_5}_\zeta \zeta^5.
\end{equation}
Where the coefficients are defined as follows
\begin{align*}
{d_1}_\zeta &= 1, \\
{d_2}_\zeta &= \frac{1}{2}(3-q_0), \\
{d_3}_\zeta &= \frac{1}{6}(11 - 5q_0 - 3{q_0}^2 - j_0),\\
{d_4}_\zeta &= \frac{1}{24}(50 - 7 j_0 - 26 q_0 + 10 j_0 q_0 + 21 q_0^2 - 15 q_0^3 +s_0),\\
{d_5}_\zeta &= \frac{1}{120} (274 - 47 j_0 + 10 j_0^2 - 154 q_0 + 90 j_0 q_0 + 141 q_0^2 - 105 j_0 q_0^2 - 135 q_0^3 + 105 q_0^4 +9s_0 - 15q_0s_0 -l_0).
\end{align*}

\section{Appendix B: BAO $D_V$ parameter series expansion}
The $D_V$ series expansion in $\zeta$ is given by the following expressions.
\begin{equation}
\label{eq:zetaDV}
{D_V} = {D_1}_\zeta \zeta + {D_2}_\zeta \zeta^2 + {D_3}_\zeta \zeta^3 + {D_4}_\zeta \zeta^4 + {D_5}_\zeta \zeta^5.
\end{equation}
Where the coefficients are defined as follows
\begin{align*}
{D_1}_\zeta &= 1, \\
{D_2}_\zeta &= \frac{1}{3}(1-2q_0), \\
{D_3}_\zeta &= \frac{1}{36}(7 - 10q_0 + 29{q_0}^2 - 10j_0), \\
{D_4}_\zeta &= \frac{1}{324}(44 - 39 j_0 - 57 q_0 + 258 j_0 q_0 + 117 q_0^2 - 376 q_0^3 + 27s_0), \\
{D_5}_\zeta &= \frac{1}{19440}(2017 - 1536 j_0 + 3540 j_0^2 - 2492 q_0 + 6990 j_0 q_0 + 4638 q_0^2 - 36300 j_0 q_0^2 - 10460 q_0^3 + 35395 q_0^4\\
&  + 702s_0 - 5400q_0s_0 - 378 l_0).
\end{align*}


\bibliographystyle{JHEP}
\bibliography{references}

\end{document}